# Radiative Cooling and Thermoregulation in the Earth's Glow

*Jyotirmoy Mandal,[1]\* Sagar Mandal,[2] John Brewer,[1] Arvind Ramachandran,[3] Aaswath Pattabhi Raman[1]\**

[1] Department of Materials Science and Engineering, University of California, Los Angeles, California, USA
[2] Independent Researcher, 523 Broadway E, Seattle, Washington, USA
[3] Department of Civil, Environmental and Sustainable Engineering, Arizona State University, Tempe, Arizona, USA

**Contact:** Jyotirmoy Mandal (jyotirmoymandal@ucla.edu) and Aaswath Raman (aaswath@ucla.edu)

## Abstract

Passive radiative cooling involves a net radiative heat loss to the cold outer space through the atmospheric transmission window in the long-wave infrared (LWIR) wavelengths. Due to its passive nature and net cooling effect, it is a promising alternative or complement to electrical cooling. For efficient radiative cooling, an unimpeded view of the sky is ideal, with prior work thus focused on roofs and sky-facing surfaces. However, much of the surface area of typical buildings are vertically oriented, with ≳50% of their field of view subtended by terrestrial features. Under sunlight, these features become warm and thermally irradiate vertical facades of buildings. Since building facades are made of materials that are broadband emitters and absorbers of thermal radiation, this heating effect can dramatically counter heat loss to the sky, diminishing or even reversing the radiative cooling process.

We show that selective LWIR emitters on vertical building facades can experience higher cooling than achievable by using conventional broadband thermal emitters. They achieve this enhancement by exploiting the differential transmittance of the atmosphere towards the sky and between terrestrial objects. Intriguingly, this effect diminishes and even reverses during the winter – representing a passive thermoregulation that arises with seasonal variation in the terrestrial thermal environment. This thermoregulation can be exhibited by modest-to-highly solar reflective (> 0.80) vertical facades under direct sunlight, and even less solar reflective facades that are diffusely lit. We experimentally demonstrate relative cooling of 0.43-0.46°C in warm weather and heating of 0-0.6°C in cold weather using a low-cost, scalable selective emitter completely exposed to the environment. We further demonstrate differential radiative power flows of up to ~40 W m$^{-2}$ through a direct heating experiment in cold weather.

We show that selective LWIR emitters can be conveniently made from common polymers and ceramics, and are applicable on a variety of building envelopes, including walls and windows. Preliminary estimates of energy savings, achievable by using LWIR rather than broadband emitters on vertical building facades, yield values similar in magnitude to savings from cool roofs. Collectively, the findings highlight a remarkable opportunity to harness untapped energy savings in buildings.

## Introduction

With global increases in temperatures posing fundamental economic, health and security risks to human civilization, maintaining habitable built environments has become one of the most important challenges of our times. Cooling and heating buildings currently consumes 12% of energy globally, with energy use for cooling in particular expected to grow dramatically by 2050.[1] Prevalent cooling methods, such as air conditioners (ACs), move heat outside interior spaces, while consuming large amounts of electricity, generating their own heat, and resulting in direct and indirect greenhouse gas emissions. Furthermore, in urban areas, the net heat from dense clusters of AC units and the prevalence of human-made structures that trap solar heat and inhibit evaporative cooling, lead to heat islands that experience even hotter temperatures. Indeed, active cooling methods may exacerbate climate change[2] and resulting cooling needs.[3–5] Therefore, they are not sustainable solutions for large-scale thermoregulation of built environments.

Controlling radiative heat flows into and out of buildings is a central mechanism by which the need for active cooling and heating can be reduced. To that end, decades of research has explored a range of strategies for controlling solar heat gain through different components of the building envelope (e.g. roofs, walls, windows, and skylights). Innovations in materials synthesis and optical design have enabled tailored responses to different components of the solar spectrum (UV, visible and near-infrared wavelengths). However, in addition to solar gain, the built environment radiatively emits and absorbs heat from its immediate environment over infrared wavelengths (λ~ 2.5-40 μm). This ubiquitous heat exchange has, in large part, not been optimized and leveraged to enhance efficiency. One important exception has been the radiative cooling of sky-facing surfaces of buildings.

Radiative cooling involves the radiation of terrestrial heat through the long-wavelength infrared (LWIR, $\lambda \sim 8 - 13\ \mu m$) atmospheric transmission window into outer space. Because the earth is at a higher temperature (~290 K) than outer space (~3 K), the radiative heat loss can be large if the surface radiating heat has a high emittance ($\epsilon$) in the LWIR wavelengths ($\epsilon_{LWIR}$).[6] If a surface has a sufficiently high solar reflectance ($R_{solar}$) it can also achieve a net heat loss and radiatively cool to sub-ambient temperatures under sunlight.[6] Crucially, radiative cooling is fundamentally passive in nature and yields a net cooling effect, making it a sustainable alternative to conventional active cooling systems. Research on radiative cooling has yielded a range of designs, ranging from traditional white paints,[7–9] porous polymers,[6,10] and silver-backed multilayer films,[11] polymers,[12–14] dielectric emitters,[15–17] and polymer-dielectric composites.[12,18–21] These designs encompass both selective thermal emitters, which are optimal for achieving deep sub-ambient temperatures, and broadband thermal emitters which are suitable for operation at or near-ambient temperatures. Other works have also identified strategies to enable radiative cooling along with thermoregulation in different weather and climate conditions.[22–24]

While radiative cooling has been well-studied for horizontal, sky-facing surfaces, the majority of the surface area of a typical building's envelope may be vertically oriented. The possibility that vertical facades such as walls can benefit from radiative cooling to the sky has, to our knowledge, remained unexplored. Unlike roofs, vertical facades have both the cold sky and the warm terrestrial environment in view. In such a scenario, the thermal glow from terrestrial features can drastically reduce, or even reverse, radiative cooling. Therefore, identifying a mechanism and materials that reduce heat gain from the terrestrial environment, while enabling heat loss to the cold sky, is crucial for the radiative cooling of vertical surfaces.

In this report, we demonstrate that scalable, selective LWIR emissive materials can optimize radiative heat flows on vertical surfaces and thereby hold the potential for substantial, untapped energy savings. Our solution is based on the differential transmittance of the atmosphere towards the sky (narrowband, LWIR) and between terrestrial objects (broadband). We first show that selectively LWIR emitting radiative coolers can reflect large bandwidths of broadband thermal radiation from the earth, even as they radiate and lose LWIR heat into the sky. Consequently, they can yield considerably greater cooling then achievable by radiative coolers with broadband thermal emittance. This is significant for buildings, as traditional construction materials[25] white paints[6,8,25] and composites[19] are broadband emitters. We experimentally demonstrate that vertically oriented selective LWIR emitters exposed to normal atmospheric convection exhibit ~0.43-0.46°C cooler temperatures than broadband thermal emitters when exposed to hot urban environments. Intriguingly, the enhanced cooling by selective LWIR emitters diminishes, or even reverses during the winter. This indicates that unlike broadband emitters that can overcool in the winter, selective LWIR emitters can optimize radiative heat flows in a seasonally varying terrestrial environment to achieve passive thermoregulation – an effect which, to our knowledge, has not been previously reported. Our analysis shows that this thermoregulation effect can persist for modest-to-highly solar reflective (> 0.80) vertical facades that are directly sunlit, and even dark solar reflective facades that are diffusely lit. We then experimentally highlight that a range of highly scalable radiative cooling designs, such as metallized plastics, paint resins and ceramics, have the spectral characteristics needed to outperform conventional broadband emitters. Collectively, our results highlight a remarkable opportunity for the built environment worldwide: immediate gains in efficiency and passive thermoregulation of vertical facades are achievable by replacing conventional broadband emissive building materials with selectively LWIR emissive designs.

**Theoretical Model**

Research on radiative cooling typically assumes a scenario where a horizontal radiative cooler radiates heat under an unobstructed view of the sky. However, this assumption neglects a large fraction of the surface area for radiative heat transfer in buildings: walls and other vertical facades. Walls have at least half of their field of view subtended by terrestrial features. Roofs, too, may have their view of the sky obstructed by taller buildings. The panoramic thermographs in Fig. 1A and S1,[26] representing different landscapes and weathers across the world, show representative examples. The presence of terrestrial objects in the field of view has two effects. Firstly, it reduces the spatial window for heat loss into the sky. Secondly, terrestrial objects themselves radiate significant amounts of heat, especially when they reach high temperatures under sunlight (e.g. > 60˚C for roads and pavements).[27,28] Effectively this replaces the heat-sink of the sky with heat sources. Consequently, the cooling potential ($P_{cooling}$) of a vertical surface, usually defined as the difference between the radiated power from the surface ($I_{emitter}$) at ambient temperature and the downwelling hemispherical atmospheric irradiance ($I_{sky}$) (Fig. 1B), now takes the form:

$$P_{cooling} = (I_{emitter} - v\, I_{sky}) - (1-v)\, I_{earth} \quad (1)$$

Here $I_{earth}$ represents the 'earth glow' or hemispherical irradiance from the earth (Fig. 1B), and the view factor $v$ is $\leq 0.5$ (SI, Section 3). The problem is further compounded by the atmosphere, which is thick (> 80 km) and appreciably transparent only in the LWIR wavelengths along skyward directions, but much thinner (~$10^0$-$10^2$ m) and transparent across the thermal spectrum between buildings and their environment (Fig. S3).[26] Consequently, while radiative heat loss to outer space occurs in the narrow LWIR band (λ~8-13 µm) (Fig. 1B and C, upper panel), radiative heat gains from terrestrial sources is broadband (λ~2.5-40 µm) (Fig. 1B and C, lower panel), i.e. both inside and outside the LWIR band.

The effect of these factors on the cooling performance of broadband emissive radiative coolers like paint coatings can be significant. As shown in Fig. 1C, since radiated heat scales as $T^4$, even moderately above-ambient terrestrial heat sources can have a broadband heating potential that counters or outweighs the LWIR cooling potential of the sky. This is also observed in horizontal pyrgeometric measurements of the thermal environment, which is hotter than the ambient air in warm locations (Table S1).[26] Given that vertical facades and many urban roofs have terrestrial objects blocking their view of the sky, and that building envelopes are broadband thermal emitters/absorbers,[19,25] this means that heat radiated from terrestrial features can greatly reduce, or even reverse, radiative cooling.

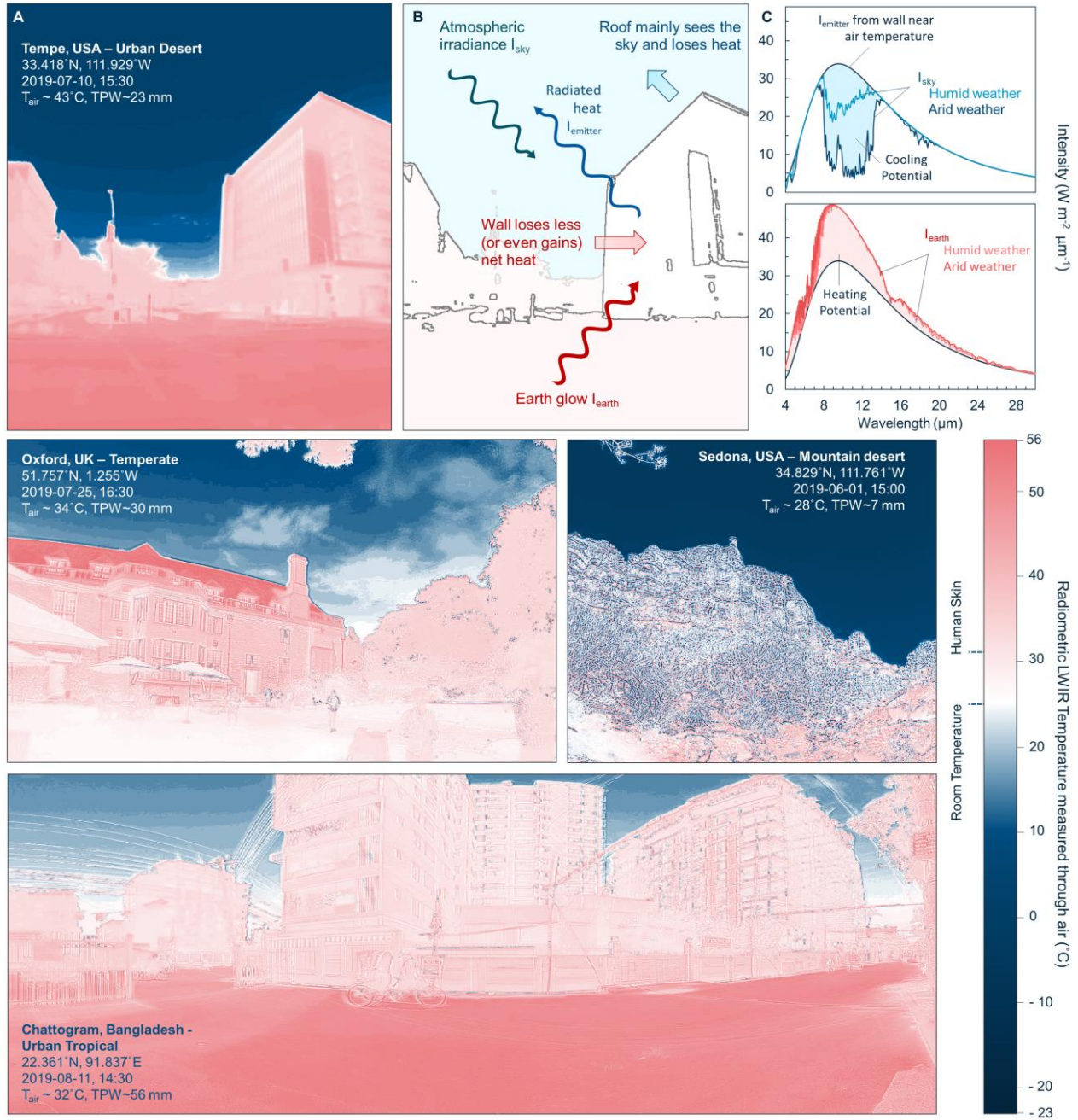

**Figure 1. A.** Panoramic LWIR thermographs of locations in Tempe and Sedona in the US, Oxford in the UK, and Chattogram in Bangladesh, representing rural and urban locations in different climate zones from the vantage point of a vertical wall. Further examples are shown in Fig. S1.[26] The colour bar represents effective radiative temperatures assuming ε=1 for the environment (SI, Section 1).[26] In all situations, the ground and other terrestrial features become considerably warm during the day, while the sky, depending on the total precipitable water (TPW) in the atmosphere, exhibits varyingly cold LWIR temperatures. It is clear from the figures that terrestrial features become strong radiators of heat in the daytime, warming objects like walls and humans in their view. **B.** Schematic showing the possible radiative heat transfer between a vertical wall and the ground and sky in its view. **C.** The radiated power ($I_{emitter}$) from a perfect broadband emitter at the ambient temperature $T_{amb} = 32°C$, and hemispherical irradiances from the ground ($I_{earth}$) at effective radiative temperature $T_{ground} = 55°C$ and the sky ($I_{sky}$) as seen through the atmosphere (Fig. S3).[26] Cases for two different humidities (low and distinctly high TPW values of 10.5 and 58.6 mm respectively) are shown. Shaded areas show the possible heat loss to the sky (blue) and gain from the ground (red).

Our solution to this problem arises from the observation that the differential transmittance of the atmosphere enables narrowband (LWIR) heat loss to the sky, and broadband heat gain from the terrestrial environment. We propose that radiative coolers which selectively emit and absorb radiation in the LWIR atmospheric window, and reflect other thermal wavelengths, can optimally harness this phenomenon to enable improvements in the net heat flows into and out of buildings. To show this, we choose the example of a vertical wall similar to that in 1B, which has equal views of the cold sky and a hot ground under the sun (Fig. S2).[26] As shown in Fig. 2A, a broadband emitter like a traditional paint coating absorbs and emits thermal radiation both within and outside the LWIR window. Consequently, for a perfect broadband emitter, the radiative loss to the sky in the LWIR (blue shaded area) is negated or outweighed by the radiative gain from the ground (red shaded area), leading to heating. However, if heat is only gained or lost in the LWIR window and reflected elsewhere (Fig. 2B), broadband heat gain can be minimized without reducing LWIR heat loss to space, resulting in cooling or significantly reduced heating compared to those achievable by broadband emissive radiative coolers.

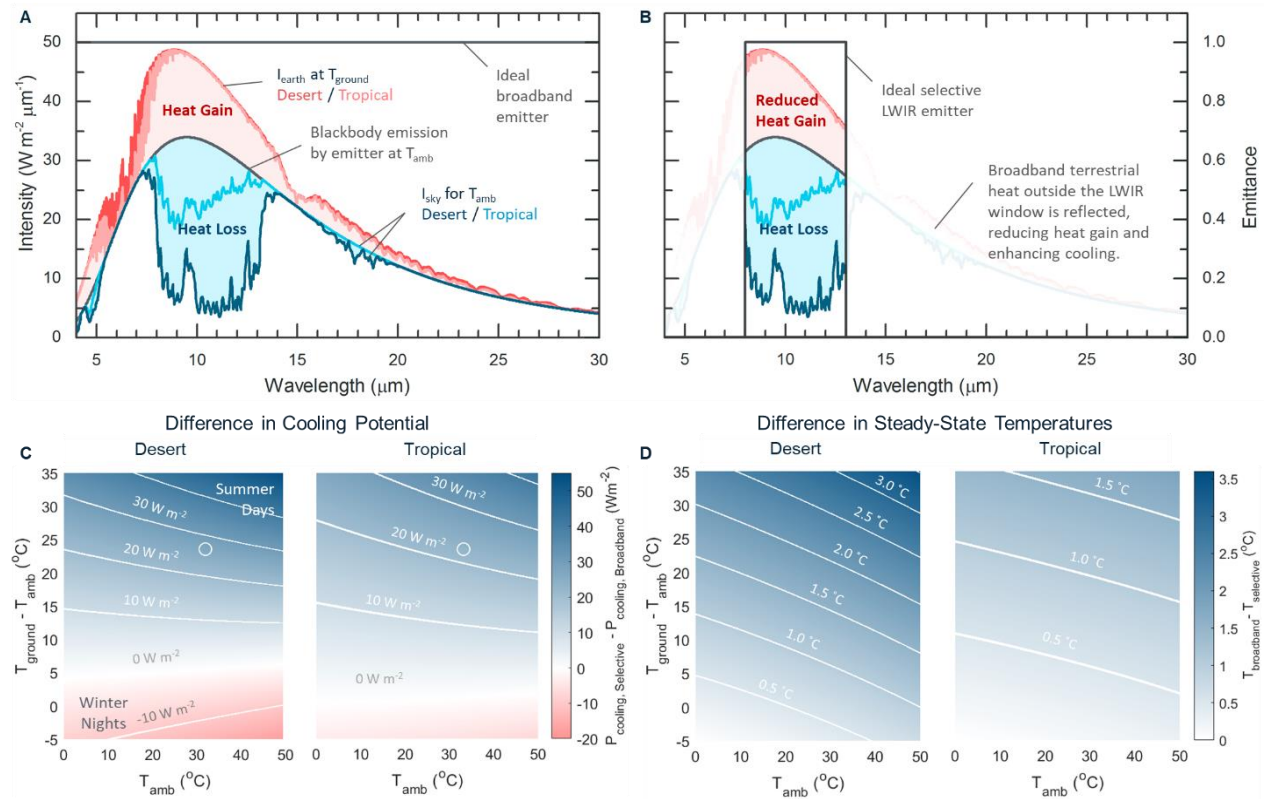

**Figure 2.** (**A**) The possible heat gain (red) from the ground (at $T_{ground}$, here 55°C) and heat loss (blue) to the sky, as shown for an ideal broadband emitter at ambient temperature $T_{amb}$ (here 32°C) under desert (TPW 10.5 mm) and distinctly humid tropical (TPW 58.6 mm) climates. Note that the irradiances $I_{sky}$ and $I_{earth}$ are hemispherical values. For a vertical emitter with equal views of the sky and the earth, the red and blue areas are weighted by ½. For more complicated environments, further irradiances and view factors can be added as an extension of this scenario. The sum of the red and blue shaded areas represent the net cooling potential for the given $T_{ground}$ and $T_{amb}$. (**B**) Analogue for an ideal selective LWIR emitter. By reflecting thermal radiation outside the window, the selective emitter filters out much of the broadband thermal radiation from the ground, and enables greater cooling when the ground in view is hot. (**C**) Differences between $P_{cooling}$ of an ideal selective LWIR emitter and broadband emitter as a function of $T_{ground}$ and $T_{amb}$. Values corresponding to the data in Figs. 2A and 2B are represented with open circles. Individual cooling potentials of the two emitters are presented in the supplementary information (Fig. S5).[26] (**D**) Analogous differences between steady state temperatures of an ideal selective LWIR emitter and broadband emitter assuming zero convective heat flows and gentle winds. Individual steady state temperatures of the two emitters are presented in the supplementary information (Fig. S6).[26] It is clear from Figs. 2C and D that a selective LWIR emitter could have considerably greater cooling potentials during hot weather, and intriguingly, a relative heating potential in cold weather - signifying a passive thermoregulation capability.

The cooling potential of a vertical surface (Eq. 1) depends on a number of factors such as temperatures of the emitter ($T_{emitter}$), ambient air ($T_{amb}$) and ground ($T_{ground}$), meteorological variables, view factors of objects in the environment, and conductive and convective coefficients ($h$) of materials. Together, these determine $I_{emitter}$, $I_{sky}$, $I_{earth}$, and non-radiative heat flows. In this study, we choose the illustrative case of a vertical wall with equal views of the ground and the sky (i.e. $v$ = 0.5) (SI, Section 3),[26] which corresponds to the minimum view factor of the terrestrial environment for a vertical surface, and is extendable to more complicated cases. Calculations of theoretical cooling potentials and steady state temperatures assuming negligible conductive heat flow show that a selective LWIR emitter can have considerable benefits. As shown in Fig. 2C, a selective LWIR emitter achieves a significantly greater $P_{cooling}$ than a broadband emitter when $T_{ground}$ is considerably greater than $T_{amb}$, as might occur on sunny summer days. In deserts, where lower total precipitable water levels enable greater LWIR heat loss into the sky, $P_{cooling}$ can be higher by more than 50 Wm$^{-2}$, which is comparable to $P_{cooling}$ of cool roofs[25] and would rise favourably for tall buildings with greater vertical surface areas. A similar effect occurs even in exceptionally humid conditions. Conversely, when $T_{ground} < T_{amb}$, a selective emitter shows a lower $P_{cooling}$, which could prevent undesirable heat loss during winters or cold nights. A similar trend is seen in the differences in steady state temperatures (Fig. 2D). Under mild winds ($h$~ 10 W m$^{-2}$ K$^{-1}$), a selective LWIR emitter stays ~3˚C cooler when the ground is hot, and at the same temperature as the broadband emitter when the ground is colder than the ambient air (Fig. S10).[26] It should be noted that the results in Fig. 2 are calculated independent of solar heat gain, i.e. assuming a $R_{solar}$ of 1 or no solar irradiance. Additional analysis shows that the summertime cooling effect between selective LWIR and broadband emitters persists for solar heat gains as high as 50-100 W m$^{-2}$, while relative wintertime heating is enhanced (SI, Section 10).[26] The persistence of this cooling effect is due to the large difference in the ambient thermal irradiance ($vI_{sky} + (1 - v) I_{earth}$) absorbed by selective and broadband emitters. Thus even modest $R_{solar}$ ~ 0.80-0.95 for directly illuminated vertical surfaces, and < 0.50 for diffusely illuminated surfaces, still yields the seasonal cooling and heating effect (SI, Section 10).[26] The relative cooling during the summer, and the diminished cooling or even heating during the winter, suggests that the capabilities of selective LWIR emitters can go beyond radiative cooling to a novel, passive thermoregulation as the terrestrial environment varies with the seasons.

**Experimental Demonstration**

To verify the theoretical results with proofs-of-concept, we performed experiments where we exposed vertically oriented selective and broadband thermal emitters to summer and wintertime environments (Fig. 3A). A 508 μm thick silvered Poly(4-methyl-1-pentene) (PMP) sheet (Fig. S7), first demonstrated as a radiative cooler by Grenier,[29] was chosen as the selective emitter, and a 127 μm thick silvered poly(vinylidene difluoride) (PVdF) sheet (Fig. S7) was chosen as the broadband emitter. The setup in Fig. 3A was exposed to the environment, facing away from direct sunlight to avoid the compounding effects of differential $R_{solar}$, in two locations, Los Angeles and Nashville (Table S2).[26] In Los Angeles, which represented summer weather, the environment was a parking lot in a semi-urban area, and the duration of exposure was ~ 40 minutes. In Nashville, which represented winter weather, the environment was an uphill view of a meadow and the duration of exposure was 1 hour. Since both samples are silvered and have similarly high $R_{solar}$ (~0.90 for PMP and ~0.93 for PVDF), differential absorption of the indirect sunlight was small (< 3 Wm$^{-2}$), making thermal irradiances the dominant radiative mode of heat transfer. Notably, the samples were completely exposed to the air, making convection in windy Los Angeles significant.

As predicted by our model, the PMP selective emitter stays 0.43˚C cooler on average over a 40-minute period relative to the PVdF broadband emitter under warm weather conditions in Los Angeles (Fig. 3B). During intervals when windspeeds are lower, as much as 0.7˚C relative cooling is observed. A similar cooling by 0.46˚C is observed in another experiment conducted nearby at a different time (SI, Section 4).[26] The results are consistent with the theoretical model (SI, Section 4).[26] The cooling effect reverses when the weather and the ground is cold, as evidenced by the PMP being 0.6˚C warmer in Nashville (Fig. 3B). The large relative warming is qualitatively consistent with our model, but does diverge from the theoretical predictions of near-zero temperature difference between the selective and ideal cases (SI, Section 4).[26] We attribute this to the cold earth occupying substantially more than half the field of view in this experiment. An additional wintertime test performed in Los Angeles, where the sky:earth view factor ratio was closer to the 1:1 assumption of our model, yielded near-zero temperature difference, as theoretically predicted. A summary of the results, compared against theoretical predictions, is presented in Fig. 3C. Accounting for uncertainties in the measurements and the theoretical predictions, the experimental and theoretical results are generally in good agreement (SI, Section 4). A near-zero temperature difference was also observed under winter conditions during the initial stage of the differential heat flow experiment described below (SI, Section 4, Fig. S11). Collectively, the results substantiate our theoretical model after accounting for environmental effects (SI, Section 4),[26] and show that selective LWIR emitters on vertical outdoor facades may not only enable greater cooling over broadband designs during the summer, but also reduce or reverse heat loss in the winter.

Since PMP and PVdF are non-ideal selective and broadband emitters respectively, and as used were not at their optimal optical thicknesses, the difference in steady-state temperatures observed appears relatively modest at first glance. However, they do substantiate the theory and represent a large difference in heat flow (SI, Section 5).[26] To demonstrate this experimentally, we heated the PMP and PVdF samples at different rates under vertical exposure to the environment during winter in Los Angeles. As shown, the selectively emissive PMP emitter stays warmer than the broadband emissive PVdF when heated at the same rate (Fig. 3D), or conversely, has a lower heat outflow for a given temperature (Fig. 3E). The differences in heat outflow, e.g. ~ 12 and 33.5 $Wm^{-2}$ when both emitters are 5 and 15°C warmer than the cold air respectively, are purely radiative and significant. More importantly, they are close to the predictions of our theoretical model (Fig. 3E. SI Section 3 and 4), and therefore indicate that the small temperature differences correspond to large differences in radiative heat flows, and thus energy savings (SI, Section 5).[26]

We emphasize that unlike in most radiative cooling experiments,[11,19] the above performances were all achieved without convection or radiation shields. Natural and forced convection from the air thus reduces the observed temperature difference, but makes the experimental conditions close to the likely real-world utilization of radiative cooling on vertical surfaces. Given the temperature drops and differences in observed heat flows, the results demonstrate the potential of selective LWIR emitters for passive thermoregulation of buildings.

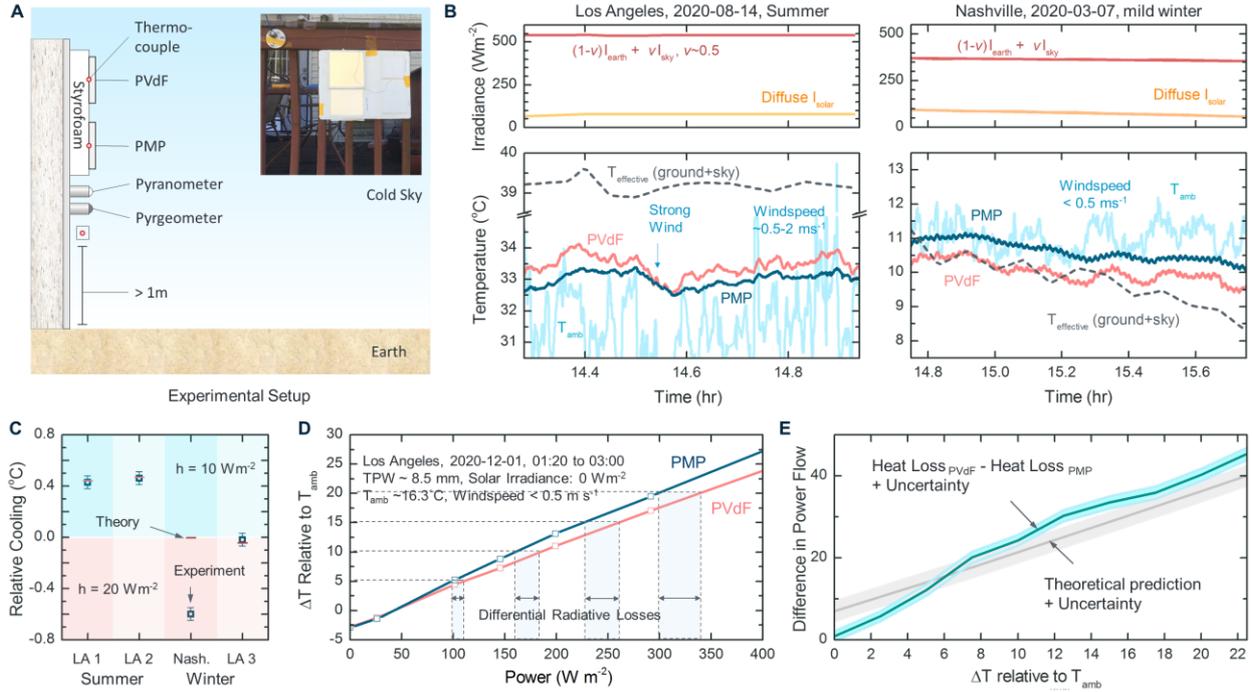

**Figure 3. (A)** Set up for experimental demonstration of the passive thermoregulation capability of selective LWIR emitters. **(B)** Solar irradiance, angle-integrated thermal irradiance, effective radiative temperature ($T_{effective}$) of a 150° field-of-view along the horizontal direction, ambient air temperature ($T_{amb}$), and temperatures of the PMP selective emitter and PVdF broadband emitter in Los Angeles and Nashville. In Los Angeles, which represents windy, summer weather, the PMP is 0.43°C cooler than the PVdF, and in Nashville, which represents calm, cold weather, the PMP is 0.6°C warmer. The results substantiate the theory (SI, Section 4),[26] and indicate that selective emitters can enable cooling in warm weather and heating in cold weather relative to broadband emitters. **(C)** Comparison of the experimentally observed (squares) and theoretically predicted (bars) steady-state temperature differences. Data for the summertime Los Angeles (LA1) test and wintertime Nashville (Nash.) test is presented in Figure 3B. The results for LA2 and LA3 tests are presented in Section 4 of the supporting information. The error bars represent 0.05°C uncertainty in the experimental measurements. **(D)** A further experiment was done in Los Angeles during the winter to demonstrate the large differential heat flows through the selectively LWIR emissive PMP and broadband emissive PVdF emitters. Under cold conditions, the selective PMP emitter exhibited warmer temperatures when subjected to the same heat flow. Shaded areas show that at same temperatures relative to $T_{amb}$, the PVdF has greater radiative heat loss than the PMP, as expected in the winter. **(E)** The differential radiative heat losses of selective PMP and broadband PVdF emitters. The PMP has a lower heat loss, e.g. by ~12 Wm$^{-2}$ when 5°C warmer than $T_{amb}$, and by ~33.5 Wm$^{-2}$ when 15°C warmer. Accounting for experimental uncertainties, and the complex environment (SI, Section 4) compared to the simplifications in our model (SI, Section 3), the results are consistent with the theory.

In addition to PMP, we also investigated a range of selective LWIR emitters, including plastics, polymer resins and ceramics, for potential use on buildings. Fig. 4 and S14[26] shows several examples, which are either made from common materials, or are already commercially available. Of these, poly(4-methyl-1-pentene) (PMP or more commonly known as TPX®), was first demonstrated as a radiative cooler by Grenier in 1979,[29] while poly(vinyl fluoride) (PVF) was demonstrated by Catalanotti in 1974.[30] We introduce three further examples – metalized polypropene (PP), biaxially oriented poly(ethene terephthalate) (BoPET, commonly known as mylar), and alumina ($Al_2O_3$) ceramic tiles, and note that thin films of common paint resins based on poly(methyl methacrylate) (PMMA) and poly(dimethylsiloxane)/silicone (PDMS), and common household materials such as scotch-tape, also exhibit selective LWIR emittance (Figs. 4, S14 and S17).[14,26] Fig. 4B shows the selectivity (quantified as $\varepsilon_{LWIR}/\varepsilon_{non-LWIR}$) of commercially available variants of these materials derived from their emittance spectra (SI, Section 1 and Fig. S14)[26]. Despite being off-the-shelf and not optimized for optical thickness, the selectivity of these materials compare favourably with those of reported selective designs in the literature - suggesting immediate suitability from an optical standpoint.

In terms of practical use, we note that these materials are used in large scales.[31,32] Indeed, some, like polypropene and mylar are sufficiently common even in their metallized forms to be sourced from plastic waste.[31] Further discussions about their scalability, and cost relative to exterior paints, are included in the supplementary information (Section 6).[26] Notably, some of these materials have long been established as radiative coolers[13,29,30,33] and recently used as components in large scale broadband emissive designs.[6,12,13,19,25] Furthermore, these materials can come in white, silvered, and (in the case of transparent conductor-backed polymers and Alumina) transparent variants (Fig. 4 and S14). These make the designs suitable for a wide range of building facades such as walls and windows. For white or silvered designs on walls, possible modes of applications include silver or white 'wallpapers' (e.g. metallized PMP, PVF, PP, Mylar and their white polyethene coated variants, SI, Section 6),[26] white tiles (e.g. $Al_2O_3$ ceramics, Fig. 4A). Metallic facades, which are intrinsically broadband reflective, could have plastics laminated, or PDMS/ PMMA painted onto them. It should be noted that while a high solar reflectance is desirable for maximizing energy savings for hot summertime scenarios, depending on the façade orientation and solar incidence, $R_{solar}$~0.80-0.95 for directly illuminated surfaces and even < 0.50 for diffusely illuminated surfaces may yield comparable savings (SI, Section 10).[26] The designs we demonstrated experimentally have $R_{solar}$~0.95 or more (SI, Section 7),[26] fall on the more desirable end of this range, and could be further optimized for $R_{solar}$. Furthermore, solar absorption-induced losses in summertime savings can be partially compensated for by reductions in heat losses in the winter (SI, Section 10).[26] For transparent facades such as windows, solar absorption is less important. Possible transparent designs include laminated or painted transparent polymers, or sputtered $Al_2O_3$, on IR reflective (low-E) glasses (Figure S15).[26] For Al2O3, which is intrinsically reflective beyond $\lambda$~13 μm, even bare glass would suffice.[34] We further note that in specific cases where diffuse solar reflection off walls is an issue, retroreflective[35] or less reflective variants of the designs above, or transparent designs, could be used (SI, Section 9). These designs can be visually similar to traditional building envelopes. Given the scalability, versatility and potential uses of selectively LWIR emissive materials, it is conceivable that they could be widely used on vertical facades.

Manuscript - Radiative Cooling and Thermoregulation in the Earth's Glow- Mandal et. al.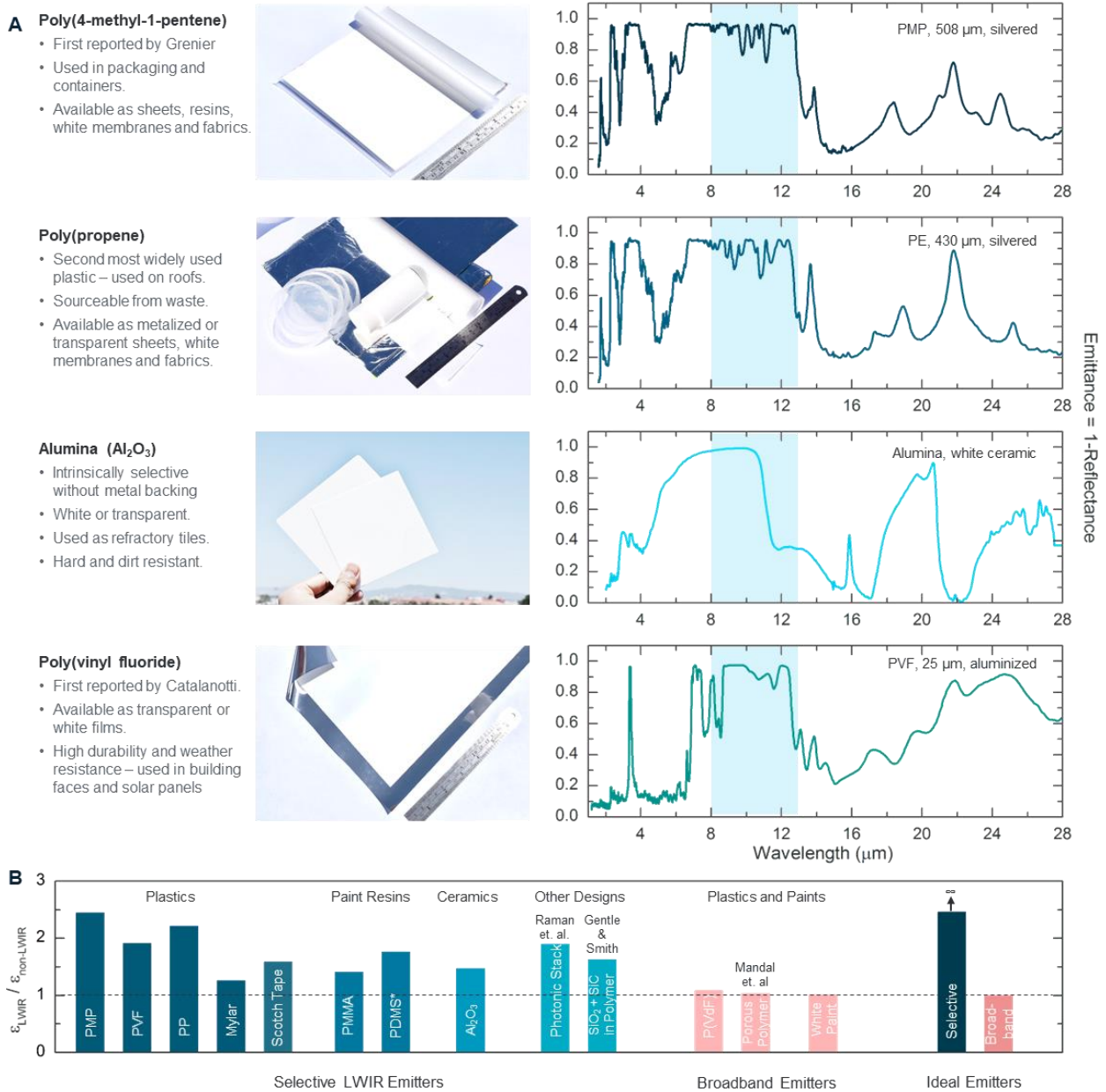

**Figure 4.** Selective LWIR emitters come in different varieties. (**A**) Common examples include poly(4-methyl-1-pentene), polypropene, alumina and poly(vinyl fluoride), and are available in various forms as shown in the pictures. Further examples are provided in the supplementary information (Section 6).[26] (**B**) The LWIR selectivity of different materials, characterized using the figure of merit $\eta = \epsilon_{LWIR}/\epsilon_{non\ LWIR}$ (SI, Section 1).[26] Note that the values of $\eta$ we demonstrate are for materials obtained off-the-shelf, and could be higher if optimized for optical thickness. * Indicates calculation from experimentally derived complex refractive index.

**Discussion**

The thermoregulation capability of LWIR emitters on vertical building facades, and the availability and variety of scalable LWIR emissive designs, could lead to significant energy savings. The precise amount depends on factors like geographic location, landscape (e.g. urban versus rural), and seasonal and diurnal variations of the atmospheric and terrestrial irradiances. As a preliminary effort, here we extend the model used in Fig. 2 to a quasi-steady state model that accounts for a fixed indoor temperature of 25˚C and vertical facades with different effective thermal conductances (U-Value accounting for thermal mass) (Fig. 5A), to make preliminary estimates of building-level energy savings (SI, Section 5).[26] Our model shows that during peak summer in the arid southwestern US or subtropical South Asia, selective LWIR emitters can achieve cooling savings of ~0.01-0.04 kWh m$^{-2}$ day$^{-1}$ for walls depending on the insulation, 0.015-0.065 kWh m$^{-2}$ day$^{-1}$ for windows depending on the glazing type, and > 0.06 kWh m$^{-2}$ day$^{-1}$ for metal sheets (Fig. 5A-B). During winter, the heating savings are ~ 0 kWh m$^{-2}$ day$^{-1}$ for walls, ~0-0.005 kWh m$^{-2}$ day$^{-1}$ for windows and > 0.012 kWh m$^{-2}$ day$^{-1}$ for metal sheets in the Southwestern US. For subtropical South Asia, heating penalties are observed, but depending on the U-value, they are ~2-10x lesser than summertime savings. For the non-ideal and unoptimized PMP emitter we used in our experiment, the savings are lower, as expected, but significant nonetheless, and in winter conditions, shows relative energy savings over the ideal case (Fig. 5B).

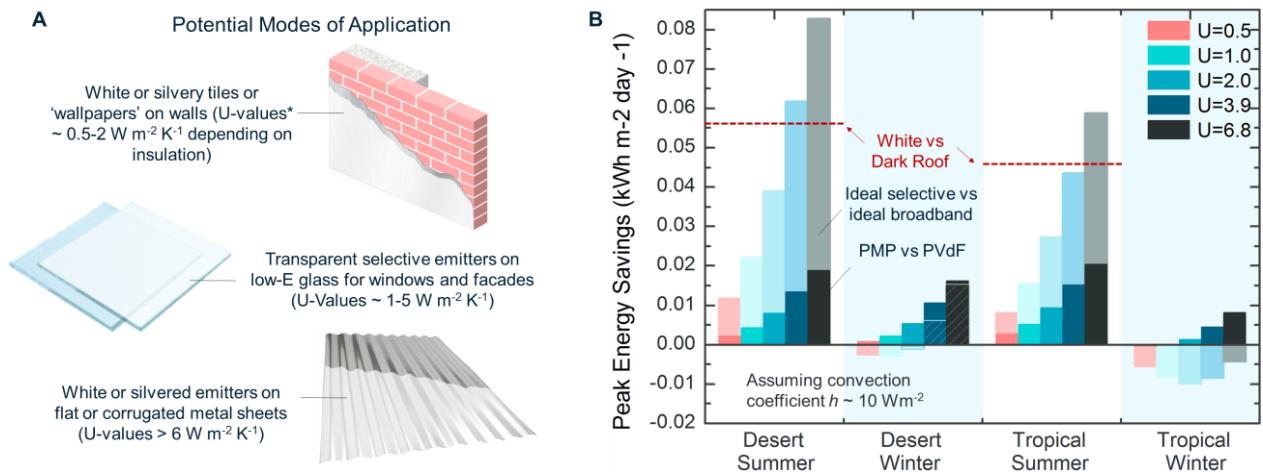

**Figure 5.** Potential modes of application of selective emitters and approximate energy savings based on a developed model (SI, Section 5).[26] (**A**) Potential ways in which selective LWIR emitters could be applied onto vertical facades made from materials with different U-values. For brick, the * indicates an effective U-value (SI, Section 5)[26] that accounts for thermal mass. For glass and metal, the thermal mass has minimal impact. (**B**) Peak summer and wintertime energy savings enabled by a selective LWIR emitter in desert and tropical locations for wall materials with different U-values (Fig. 5A). A gentle wind, corresponding to a convection coefficient of 10 W m$^{-2}$ K$^{-1}$, was assumed. Desert location corresponds to Palm Springs in the Southwestern US, while the tropical location corresponds to Kolkata in South Asia. Details of the model used is provided in Section 5 of the supplementary information.[26] The dotted lines indicating savings for white roofs on single-family houses during the summertime are derived from raw data for Phoenix (desert) and Miami (tropical) from a work by Baniassadi et. al.[36]

We note here that these preliminary results are derived using a first-order model to motivate detailed studies by the building energy researchers. Nonetheless, it is notable that they are similar in magnitude to the savings achievable by white roofs that minimize solar heating (Fig. 5B),[36,37] even without adjusting for increased vertical facade area with building height. We note that for typical buildings, the area of vertical facades can be anywhere between ~ 1-10x that of roof area. If the ratio is factored into our model, then the potential peak savings of selective LWIR emitters become comparable to, or higher, than those of cool roof coatings in the summer, even with non-ideal LWIR emitters. For instance, for a tall building with brick walls (U-value ~ 2 W m$^{-2}$ K$^{-1}$) that have 10x the area of its roof, the preliminary savings for PMP vs PVdF emitters we used could be ~ 0.075 kWh m$^{-2}$ day$^{-1}$ in peak desert summer conditions, higher than the 0.056 kWh m$^{-2}$ day$^{-1}$ achievable by painting roofs white. Our findings make selective LWIR emitters worthy of exploration as energy-saving envelopes for a wide variety of vertical facades, and a complement to previously reported super-white roof coatings and other sky-facing radiative cooling designs.[6,13,19,25] As implied above, a natural extension from existing, solar-reflective cool-walls[35,38,39] (SI, Section 9) could be silvery or white LWIR emitters on brick or sheet-metal walls that offer thermoregulation capabilities with the benefits of high solar reflectance (SI, Section 7). Indeed, the leeway on $R_{solar}$ (SI, Section 10) could even allow the use of solar-infrared reflective and/or fluorescent 'cool colours', an intriguing possibility in terms of aesthetics.[40–43] The other major possibility, which can impact windows and steel-and-glass architectures (SI, Section 9) is solar transparent LWIR emissive polymers on low-E glasses or far-infrared reflective ceramics on bare glass. Paints for cooling building facades constitutes a US$ 18 billion market,[44] while low-emittance glass panels account for US$ 28 billion.[45] Corrugated metal sheets, meanwhile, are frequently used as walls of houses in developing countries (SI, Section 9). Selective LWIR emitters, therefore, hold the potential to enable large energy savings if used as vertical building facades.

In addition to use on vertical facades of buildings, selective LWIR emitters could also be used on roofs or in water cooling panels[46] in urban settings, where views of the sky are often blocked by taller buildings. The use on building facades could also be extended to vehicles, a significant fraction of whose surfaces are vertical. Another promising use could be in radiative cooling textiles. As evident from the thermal environment around the human beings in Fig. 1A, textiles like broadband emissive cotton or IR-transparent polyethene (which exposes the broadband emissive skin underneath) can result in a net heat gain (Fig. S19),[26] particularly in urban environments. Selectively LWIR emissive textiles, which could potentially be formed by metallizing PP based fabrics (Fig. S14 and S20),[26] offer a major advance in this regard. A final intriguing possibility to explore would be integration with phase-change-materials or fluidic designs,[23,24] which could amplify the thermoregulation capability of LWIR emitters.

Since Trombe's seminal study of radiative cooling,[47] subsequent works have primarily focused on LWIR heat loss through the atmosphere towards the sky. The realization that the atmosphere allows broadband radiative heat transfer between terrestrial objects, and that vertically oriented selective LWIR emitters minimize broadband heat gain from or loss to the earth while facilitating heat loss to the sky, can have significant impact and thermoregulatory applications beyond those of sky-facing radiative cooling designs. Our findings open new opportunities for selection of materials for vertical building envelopes that complement cool-roofs, and designs enabling energy efficiency. We hope that this work will spur further research in this field.

### Acknowledgments

We would like to thank Prof. David Sailor and Jyothis Anand Jayaprabha of Arizona State University for kindly providing and elucidating the raw data for reference,[36] and Dr. Tiphaine Galy and Prof. Laurent Pilon of University of California, Los Angeles, for their assistance with preliminary measurements. We would also like to thank Prof. Sir Keith Burnett for his valuable advice, and Dr. Adam Overvig and Dr. Kamal Krishna Mandal for helpful discussions.

### Funding

Jyotirmoy Mandal was supported by Schmidt Science Fellows, in partnership with the Rhodes Trust. We also acknowledge support from the Alfred P. Sloan Foundation.

### Author Contributions

J.M. originated the concept and explored materials for use as LWIR emitters. A.P.R. supervised the study. J.M. and A.P.R. designed the experiments. J.M., S.M. and A.R. performed thermography, pyrgeometry and image processing. J.B. prepared samples for experiments. J.M. and J.B. performed the outdoor experiments. J.M performed the spectroscopy and modelling. J.M., A.P.R., and S.M. wrote the manuscript.